\documentclass[%twocolumn,
showpacs,preprintnumbers,amsmath,amssymb]{revtex4}
\usepackage{amsmath}
\usepackage{graphicx}
\usepackage{subfig}
\usepackage{hyperref}
\usepackage{amssymb}
\usepackage[english]{babel}
\usepackage{epsfig}
\usepackage{wasysym}
\usepackage{graphicx}
\usepackage{indentfirst}
\usepackage{amsmath}
\usepackage{amsfonts}
\usepackage{amssymb}
\usepackage{enumerate}

\begin{document}

\newcommand{\vecbo}[1]{\mbox{\boldmath $#1$}}
\newtheorem{theorem}{Theorem}
\newtheorem{acknowledgement}[theorem]{Acknowledgement}
\newtheorem{algorithm}[theorem]{Algorithm}
\newtheorem{axiom}[theorem]{Axiom}
\newtheorem{claim}[theorem]{Claim}
\newtheorem{conclusion}[theorem]{Conclusion}
\newtheorem{condition}[theorem]{Condition}
\newtheorem{conjecture}[theorem]{Conjecture}
\newtheorem{corollary}[theorem]{Corollary}
\newtheorem{criterion}[theorem]{Criterion}
\newtheorem{definition}[theorem]{Definition}
\newtheorem{example}[theorem]{Example}
\newtheorem{exercise}[theorem]{Exercise}
\newtheorem{lemma}[theorem]{Lemma}
\newtheorem{notation}[theorem]{Notation}
\newtheorem{problem}[theorem]{Problem}
\newtheorem{proposition}[theorem]{Proposition}
\newtheorem{remark}[theorem]{Remark}
\newtheorem{solution}[theorem]{Solution}
\newtheorem{summary}[theorem]{Summary}
\newenvironment{proof}[1][Proof]{\textbf{#1.} }{\ \rule{0.5em}{0.5em}}
\hypersetup{colorlinks,citecolor=green,filecolor=magenta,linkcolor=red,urlcolor=cyan,pdftex}

\newcommand{\be}{\begin{equation}}
\newcommand{\ee}{\end{equation}}
\newcommand{\bea}{\begin{eqnarray}}
\newcommand{\eea}{\end{eqnarray}}
\newcommand{\beaa}{\begin{eqnarray*}}
\newcommand{\eeaa}{\end{eqnarray*}}
\newcommand{\Lhat}{\widehat{\mathcal{L}}}
\newcommand{\nn}{\nonumber \\}
\newcommand{\e}{{\rm e}}

%%%%%%%%%%%%%%%%%%%%%%%%%%%%%%%%%%%%%%%%%%%%%%%%%%%%%%%%%%%%%%%%%%%%%%%%

%%%%%%%%%%%%%%%%%%%%%%%%%%%%%%%%%%%%%%%%%%%%%%%%%%%%%%%%%%%%%%%%%%%%%%%%

\title{ 
Geodesic Deviation Equation in $ f (R,T)$  Gravity
}
%%%%%%%%%%%%%%%%%%%%%%%%%%%%
 \author{E. H. Baffou$^{(a)}$\footnote{e-mail:baffouh.etienne@yahoo.fr},  M. J. S. Houndjo$^{(a,b)}$\footnote{e-mail:
  sthoundjo@yahoo.fr}, M. E. Rodrigues$^{(b)}$\footnote{e-mail: esialg@gmail.com}, A. V. Kpadonou $^{(a,c)}‡$\footnote{e-mail: vkpadonou@gmail.com}, and J. Tossa$^{(a)}$\footnote{e-mail: joel.tossa@imsp-uac.org} }
%  %%%%%%%%%%%%%%%
  \affiliation{$^a$ \, Institut de Math\'{e}matiques et de Sciences Physiques (IMSP), 01 BP 613,  Porto-Novo, B\'{e}nin\\
$^{b}$\, Facult\'e des Sciences et Techniques de Natitingou - Universit\'e de Parakou - B\'enin \\
$^{c}$\, Ecole Normale Sup\'erieure de Natitingou - Universit\'e de Parakou - B\'enin\\ 
$^{d}$ \, Faculdade de F\'isica, PPGF, Universidade Federal do Par\'a, 66075-110, Bel\'em, Par\'a, Brazil.\\  
$^{e}$\,Faculdade de Ci\^encias Exatas e Tecnologia, Universidade Federal do Par\'a - Campus Universit\'ario 
de Abaetetuba, CEP 68440-000, Abaetetuba, Par\'a, Brazil }  

%%%%%%%%%%%%%%%%%%%%%%%%%%%%%%%%%%%%%%%%%%%%%%%%%%%%%%%%%%%%%%%%%%%%%%%%%%%%%%%%%%%%%%%%%%%%%%%%%%%%%%%%%%%%%%%%%%%%%%%%%%%%%%%
\begin{abstract}
\vspace{0.2cm}
In this paper, we investigate the modified Geodesic Deviation Equation (GDE) in the framework of $f(R,T)$ theory of gravity where
$R$ and $T$ are the curvature scalar and the trace of the energy-momentum tensor, respectively, using the FLRW background. 
In this way, we obtain the GR equivalent  (GDE) in  $f(R,T)$ metric formalism. We also extend our work to the generalization of the Mattig relation and perform the numerical analysis  with GDE for null vector.

\end{abstract}
%%%%%%%%%%%%%%%%%%%%%%%%%%%%%%%%%%%%%%%%%%%%%%%%%%%%%%%%%%%%%%%%%%%%%%%%%%%%%%%%%%%%%%%%%%%%%%%%%%%%%%%%%%%%%%%%%%%%%%%%%%%%%%%%%
%%%%%%%%%%%%%%%%%
%%%%%%%%%%%%%%%%%
\maketitle 
%%%%%%%%%%%%%%%%%%%%%%%%%%%%%%%%%%%%%%%%%%%%%%%%%%%%%%%%%%%%%%%%%%%%%%%%%%%%%%%%%%%%%%%%%%%%%%%%%%%%%%%%%%%%%%%%%%%%%%%%%%%%%%%%%%%%%%%%%%%%%%%%%%%%%%%
\section{Introduction}

General Relativity (GR) is the physical theory of gravity formulated by Einstein in $1916$.
It is based on the equivalence principle of gravitation and inertia, which establishes a fundamental connection between the gravitational field and the geometry of the spacetime, and on the principle of general covariance.
GR is the most widely accepted gravity theory and it has been tested in several field strength regimes, but not the only relativistic
theory of gravity \cite{eti29}.
This theory explains the gravity as the curvature of spacetime. 
The relative motion of test particles is one of the most important sources of information about the gravitational field and
spacetime geometry. This motion is described by the Geodesic Deviation Equation. It may be claimed that the GDE is one of the most important equations in relativity, as this is how one measures spacetime curvature. This aspect has been discussed by Szekeres \cite{eti25}. The spacetime curvature described by the Riemann tensor  manifests through the Geodesic Deviation Equation\cite{eti20}-\cite{eti22}, the relative acceleration of the test particles, also known as tidal acceleration.
The signification  of the equation of geodesic deviation in relation to the relative acceleration and tidal forces between two neighboring
freely falling test particles under gravity has been repeately stressed since the mid fifties.
The importance of the geodesic deviation equation (GDE) for spinless particles is studied by the authors \cite{eti23},\cite{eti24}, and they certify
its importance when we study gravitational wave phenomena and their detection.
The geodesic deviation equation (GDE) provides an elegant tool to investigate the timelike, null
and spacelike structure of spacetime geometries and the important Raychaudhuri equation \cite{eti26}, the Mattig relation \cite{eti27} and the Pirani 
equation\cite{eti28} may be obtained by solving it. 
 Recent observations of the supernovae type Ia (SNe Ia) \cite{eti15}, the cosmic microwave background radiation
(CMBR)\cite{eti16}, the baryon acoustic oscillation (BAO) surveys \cite{eti17}, the large scale structure\cite{eti18} and the weak lensing \cite{eti19},
clearly indicate that the Universe is currently expanding with an accelerating rate.
This accelerating expansion is one of the most important puzzles of contemporary physics.
As a consequence, all observations related to gravity should be described within a framework including the cosmological constant
in the Einstein field equation. Another way proposed to explain this acceleration of the Universe, is to modify the Einstein Lagrangian i.e., modified 
gravity theory known as the generalization of the Einstein field equations \cite{eti30},\cite{eti31}. Nowdays, within these
extended theories we have one a class of modified gravity theories in which the gravitational action contains a general
function $f(R,T)$. This theory namely $f(R,T)$ modified gravity, is a generalization of $f(R)$ theory of gravity and several
important results have been found in such theory \cite{eti1}-\cite{eti34}. Because of the interesting results and the progress
realized in this framework it seems there is still room to study some motivating gravitational and cosmological aspects of $ f(R,T)$ gravity which have not yet been
studied. The GDE has been studied in $f(R)$ gravity theory \cite{eti35},\cite{eti36} and the $f(T)$ gravity \cite{eti37}, where $R$, $T$ denote the curvature scalar and the torsion scalar, respectively 
and to the key with excellent results. As these authors, the goal in this paper is to extend the study of GDE in the metric context of $f(R,T)$ gravity to obtain
the GR equivalent GDE of $f(R,T)$ gravity and studying some particular cases.

%%%%%%%%%%%%%%%%%%%%%%%%%%%%%%%%%%%%%%%%%%%%%%%%%%%%%%%%%%%%%%%%%%%%%%%%%%%%%%%%%%%%%%%%%%%%%%%%%%%%%%%%%%%%%%%%%%%%%%%%%%%%%%%%%% % % % % % % % % % % % % % % % % % % % % % % % % % % % % % % % % % % % % % % % % % % % % % % % % % % % % % % % % % % % % % % % % % % % % % % % % % % % % % %  % % % % % % % % % % % % % % % % % % % % % % % % % % % % % % % % % % % % % % % % % % % % % % % % % % % % % % % % % % % % % % 
\section{ Background $f(R,T)$ cosmology}
In this section we briefly review $f(R,T)$ gravity and we provide the background cosmological equations in a
universe governed by such a modified gravitational sector. The action for this theory coupled with matter Lagrangian $\mathcal{L}_m$
is given by \cite{eti1}, 
\begin{eqnarray}
S =  \int \sqrt{-g} dx^{4} \Big[\frac{1}{2\kappa} f(R,T)+\mathcal{L}_m \Big]\,\,, \label{1}
\end{eqnarray}
where $f(R,T)$ is an arbitrary function of the curvature scalar $R$, and of the trace $T$ of the energy-momentum tensor, respectively, and $\kappa=8\pi \mathcal{G}$, $\mathcal{G}$ 
being the gravitation constant.\par
We define the energy-momentum tensor associated to the matter as
\begin{eqnarray}
T_{\mu\nu}=-\frac{2}{\sqrt{-g}}\frac{\delta\left(\sqrt{-g}\mathcal{L}_m\right)}{\delta g^{\mu\nu}},\label{2}
\end{eqnarray} 
and its trace by $T= g^{\mu\nu}T_{\mu\nu}$.
By assuming that the matter Lagrangian density $ \mathcal{L}_m $ only depends on the components of the metric tensor $ g_{\mu\nu}$, and not on its derivatives,
we obtain  
\begin{eqnarray}
T_{\mu\nu} = g_{\mu\nu}L_{m}-\frac{{2}{\partial{L_{m}}}}{\partial{g^{\mu\nu}}}. \label{3}
\end{eqnarray}
Within the metric formalism, varying the action $(\ref{1})$ with respect to the metric, one obtains the following field equations,
\begin{eqnarray}
f'(R)R_{\mu\nu}-\frac{1}{2} g_{\mu\nu}f(R,T)+(g_{\mu\nu}\Box-\nabla_{\mu}\nabla_{\nu})f'(R)= \kappa T_{\mu\nu}-
f'(T)(T_{\mu\nu}+\Theta_{\mu\nu})\,,\label{4}
\end{eqnarray}
where $f'(R), f'(T)$  denote derivates of $ f(R,T) $ with respect
to the $ R $, $T $ respectively, $\Box = g^{\mu\nu}\nabla_{\mu}\nabla_{\nu}$ is the d'Alembert operator,$\nabla_{\mu}$ is the covariant
derivative associated with the Levi-Civita connection of the metric tensor and  $\Theta_{\mu\nu}$ is determined by 
\begin{eqnarray}
\Theta_{\mu\nu}\equiv g^{\alpha\beta}\frac{\delta T_{\alpha \beta}}{\delta g^{\mu\nu}}=-2T_{\mu\nu}+g_{\mu\nu}\mathcal{L}_m
-2g^{\alpha\beta}\frac{\partial^2 \mathcal{L}_m}{\partial g^{\mu\nu}\partial g^{\alpha \beta}}\label{5}.
\end{eqnarray}
However in the present study, we assume that the whole content of the universe is a perfect fluid, and 
in this way, the  simple setting of the  matter Lagrangian  is taking $ L_{m} = -p$. Then, with the use of Eq (\ref{5}), we obtain for the variation of stress-energy of perfect fluid, the following expression 
\begin{eqnarray}
\Theta_{\mu\nu}= -2T_{\mu\nu} -p g_{\mu\nu}.\label{7}
\end{eqnarray}
Thus the equation motion (\ref{4}) becomes,
\begin{eqnarray}
f'(R)R_{\mu\nu}-\frac{1}{2} g_{\mu\nu}f(R,T)+(g_{\mu\nu}\Box-\nabla_{\mu}\nabla_{\nu})f'(R)= \kappa T_{\mu\nu}+
f'(T) (T_{\mu\nu}+p g_{\mu\nu})\,.\label{7a} 
\end{eqnarray}
By contracting the last equation by $g^{\mu\nu}$, we obtain the relation between the Ricci scalar $R$ and the trace $T$ of the 
energy-momentum tensor
\begin{eqnarray}
 f'(R)R-2f(R,T)+3\Box f'(R)=\kappa T+f'(T) (T+4p).\label{7b}
\end{eqnarray}
From the equation (\ref{7a}), on write the Ricci tensor as
\begin{eqnarray}
R_{\mu\nu}=\frac{1}{f'(R)}= \bigg[\kappa T_{\mu\nu}+\frac{1}{2}g_{\mu\nu}+\mathcal{D}_{\mu\nu}f'(R) +f'(T)(T_{\mu\nu}+pg_{\mu\nu})\bigg], \label{7c} 
\end{eqnarray}
where we defined the operator $\mathcal{D}_{\mu\nu}$ as,
\begin{eqnarray}
\mathcal{D}_{\mu\nu}=\nabla_{\mu}\nabla_{\nu}-g_{\mu\nu}\square.\label{7bb} 
\end{eqnarray}
We obtain also from the equation (\ref{7b}) the Ricci scalar defined as
\begin{eqnarray}
R =\frac{1}{f'(R)} \bigg[\kappa T+2f(R,T)-3\Box f'(R)+ f'(T)(T+4p)\bigg].\label{7d}  
\end{eqnarray}
The field equations (\ref{7a}) can be expressed as the Einstein field equations with an effective energy-momentum tensor \cite{eti3} in the form 
\begin{eqnarray}
 R_{\mu\nu}-\frac{1}{2}Rg_{\mu\nu} &=& \frac{\kappa T_{\mu\nu}}{f'(R)}  +\frac{1}{f'(R)} \bigg[\frac{1}{2}g_{\mu\nu}(f(R,T)-Rf'(R))+\mathcal{D}_{\mu\nu}f'(R)+f'(T)(T_{\mu\nu}+pg_{\mu\nu})\bigg],\nonumber \\
 &=& \frac{\kappa}{f'(R)} \bigl(T_{\mu\nu}+T_{\mu\nu}^{eff}\bigr), 
 \label{8}
\end{eqnarray} 
where
\begin{eqnarray}
 T^{eff}_{\mu\nu} = \frac{1}{k} \bigg [\frac{1}{2} g_{\mu\nu}(f(R,T)-Rf'(R))+ f'(T)( T_{\mu\nu}+p g_{\mu\nu})+\mathcal{D}_{\mu\nu}f'(R)\bigg], \label{9}
\end{eqnarray}
representing the effective energy-momentum tensor, which could be interpreted as the energy momentum tensor of the curvature fluid.\par
Here we consider the homogeneous and isotropic flat Friedmann-Robertson-Walker (FLRW) metric, where the line element is defined as
\begin{eqnarray}
 ds^{2}= -dt^{2}+a(t)^{2}\bigg[\frac{dr^2}{1-kr^2}+ r^2{d\theta}^{2}+r^2\sin^2{\theta} {d\phi}^{2}\bigg],\label{10}
\end{eqnarray}
being $a(t)$ the scale factor and $k$ the spatial curvature of the universe and the energy-momentum tensor of the matter is given by
\begin{eqnarray}
T_{\alpha\beta}  = (\rho + p)u_{\alpha}u_{\beta} + p g_{\alpha\beta},\label{10a}
\end{eqnarray}
$\rho$ and $p$ denote the energy density and the pressure of fluid, respectively.
The trace of the energy-momentum tensor (\ref{10a}) is
\begin{equation}
T = 3p-\rho.\label{10b}
\end{equation}
Thus, the Ricci scalar in this background is given by \cite{eti11}
\begin{eqnarray}
R= +6\bigg( 2H^2+\dot{H}+\frac{\kappa}{a^2(t)}\bigg). \label{11}
\end{eqnarray}

\section{Geodesic Deviation Equation IN GR}
We describe in this section a several notions about the Geodesic Deviation Equation in general relativity (GR). For this purpose, we consider two neighbor geodesics 
$C_1$ and $C_2$ with an affine parameter $\nu$ on 2-surface S (see Fig.1). The parametric equation of the surface is
given by $ x^{\alpha}(\nu,s)$ in which $s$ is the labels geodesics.
The vector field $V^{\alpha} = \frac{dx^{\alpha}}{d\nu}$ is tangent to the geodesic. 
The family $s$ has $\eta^{\alpha}= \frac{dx^{\alpha}}{ds}$ like it's tangent vector field.\\
By first considering $\mathcal{L}_{V}\eta^{\alpha}$=$\mathcal{L}_{\eta}V^{\alpha}$ $({[V,\eta]}^{\alpha} =0)$ which leads to 
$\nabla_{V}\nabla_{V}\eta^{\alpha}$ = $\nabla_{V}\nabla_{\eta}V^{\alpha}$ and so using 
$\nabla_{X}\nabla_{Y}Z^{\alpha}$-$\nabla_{Y}\nabla_{X}Z^{\alpha}$- $\nabla_{[X,Y]}Z^{\alpha}$= $ R^{\alpha}_{\beta\gamma\delta}Z^{\beta}X^{\gamma}Y^{\delta}$ 
in which $ Y^{\alpha}=\eta^{\alpha}$ and if we set $X^{\alpha}=Z^{\alpha}=V^{\alpha}$, we obtain the promised equation of geodesic
deviation ie the acceleration for this vector field as follows  \cite{eti4}-\cite{eti5}

\begin{figure}[h]
 \centering
 \begin{tabular}{rl}
 \includegraphics[width=5cm, height=5cm]{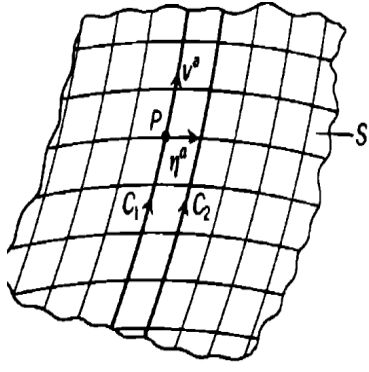}
 \end{tabular}
 \caption{Geodesic Deviation}
 \label{fig1}
 \end{figure}

\begin{eqnarray}
\frac{D^2 \eta^{\alpha}}{D \nu^2} = - R_{\beta\gamma\delta}^{\alpha}V^{\beta}\eta^{\gamma}V^{\delta}. \label{12}
\end{eqnarray}
In order to describe briefly the Geodesic Deviation Equation in General Relativity (GR), we take account the energy momentum
tensor in the form of a perfect fluid (\ref{10a}). In this way, we obtain the Einstein field equations in GR (with cosmological constant)
\begin{eqnarray}
R_{\alpha\beta} - \frac{1}{2}R g_{\alpha\beta} + \Lambda g_{\alpha\beta} = \kappa T_{\alpha\beta}. \label{13} 
\end{eqnarray}
From the Eq.(\ref{10a}), we found the Ricci scalar and Ricci tensor as follows
\begin{eqnarray}
R = \kappa(\rho - 3p) + 4\Lambda,
\end{eqnarray}
\begin{eqnarray}
R_{\alpha\beta} = \kappa(\rho + p)u_{\alpha}u_{\beta} + \frac{1}{2}\bigl[\kappa(\rho-p) + 2\Lambda\bigr]g_{\alpha\beta}.
\end{eqnarray}
Using these expressions and the Riemann tensor expressed as follows \cite{eti4}
\begin{eqnarray}
R_{\alpha\beta\gamma\delta} = C_{\alpha\beta\gamma\delta} + \frac{1}{2}\bigl(g_{\alpha\gamma}R_{\delta\beta} - g_{\alpha\delta}R_{\gamma\beta} + g_{\beta\delta}R_{\gamma\alpha}
-g_{\beta\gamma}R_{\delta\alpha} \bigr)-\frac{R}{6}\bigl(g_{\alpha\gamma}g_{\delta\beta} - g_{\alpha\delta}g_{\gamma\beta}\bigr),
\label{14}
\end{eqnarray}
where $C_{\alpha\beta\gamma\delta}$ is the Weyl tensor, the right hand side of GDE (\ref{12}) is written as 
\begin{eqnarray}
R_{\beta\gamma\delta}^{\alpha}V^{\beta}\eta^{\gamma}V^{\delta} = \biggl[\frac{1}{3}(\kappa \rho + \Lambda)\epsilon + \frac{1}{2}\kappa(\rho + p)E^2\biggr]\eta^{\alpha},
\label{15}
\end{eqnarray}

where $\epsilon = V^{\alpha}V_{\alpha}$ and $E = - V_{\alpha}u^{\alpha}$. Note that if $\epsilon = +1, 0, -1$  the geodesics are
spacelike, null or timelike, respectively. This equation is known as \textit{Pirani equation}\cite{eti6}. Well known results of the Pirani equation
 including some solutions for spacelike, timelike and null congruences have widely studied in \cite{eti7}. In our study, we generalized these
 results from the $f(R,T)$ metric formalism.
 
\section{Geodesic Deviation Equation in $f(R,T)$ Gravity}
In order to obtain the Geodesic Deviation Equation in $f(R,T)$ gravity, we give firstly the expressions for the Ricci tensor and the Ricci 
scalar (\ref{7c}) and (\ref{7d}) respectively. By employing the Riemann tensor (\ref{14}), on gets 

\begin{multline}
R_{\alpha\beta\gamma\delta} = C_{\alpha\beta\gamma\delta} + \frac{1}{2f'(R)}\Biggl[\kappa(T_{\delta\beta}g_{\alpha\gamma}-T_{\gamma\beta}g_{\alpha\delta} + T_{\gamma\alpha}g_{\beta\delta}-T_{\delta\alpha}g_{\beta\gamma}) + f(R,T)\bigl(g_{\alpha\gamma}g_{\delta\beta} - g_{\alpha \delta}g_{\gamma\beta}\bigr)\\
+  \bigl(g_{\alpha\gamma}\mathcal{D}_{\delta\beta}
 - g_{\alpha\delta}\mathcal{D}_{\gamma\beta} + g_{\beta\delta}\mathcal{D}_{\gamma\alpha}- g_{\beta\gamma}\mathcal{D}_{\delta\alpha}\bigr)f'(R)+
 f'(T)\biggl( (T_{\delta\beta}+pg_{\delta\beta})g_{\alpha\gamma}-(T_{\gamma\beta}+pg_{\gamma\beta})g_{\alpha\delta}+(T_{\gamma\alpha}+pg_{\gamma\alpha})g_{\beta\delta}-(T_{\delta\alpha}+pg_{\delta\alpha})g_{\beta\gamma}\biggr) \Biggr] \\
- \frac{1}{6f'(R)}\biggl(\kappa T + 2f(R,T) -3\square f'(R) +f'(T)(T+4p)\biggr)\bigl(g_{\alpha\gamma}g_{\delta\beta} - g_{\alpha \delta}g_{\gamma\beta}\bigr). 
\label{16}
\end{multline}
By contracting the Riemann tensor with $V^{\beta}\eta^{\gamma}V^{\delta}$ after rising its first indice, Eq (\ref{16}) becomes 
\begin{multline}
R_{\beta\gamma\delta}^{\alpha}V^{\beta}\eta^{\gamma}V^{\delta} = C_{\beta\gamma\delta}^{\alpha}V^{\beta}\eta^{\gamma}V^{\delta} + \frac{1}{2f'(R)}\Biggl[\kappa(T_{\delta\beta}\delta_{\gamma}^{\alpha}-T_{\gamma\beta}\delta_{\delta}^{\alpha} + T_{\gamma}^{\, \, \alpha}g_{\beta\delta}-T_{\delta}^{\, \, \alpha}g_{\beta\gamma}) + f(R,T)\bigl(\delta_{\gamma}^{\alpha}g_{\delta\beta} - \delta_{\delta}^{\alpha}g_{\gamma\beta}\bigr)\\
+  \bigl(\delta_{\gamma}^{\alpha}\mathcal{D}_{\delta\beta}
 - \delta_{\delta}^{\alpha}\mathcal{D}_{\gamma\beta} + g_{\beta\delta}\mathcal{D}_{\gamma}^{\, \, \alpha}- g_{\beta\gamma}\mathcal{D}_{\delta}^{\, \, \alpha}\bigr)f'(R)\\
 +f'(T)\biggl( (T_{\delta\beta}+pg_{\delta\beta})\delta_{\gamma}^{\alpha}-(T_{\gamma\beta}+pg_{\gamma\beta})\delta_{\delta}^{\alpha}+(T_{\gamma}^{\alpha}+p\delta_{\gamma}^{\alpha})g_{\beta\delta}-(T_{\delta}^{\alpha}+p{\delta}_{\delta}^{\alpha})g_{\beta\gamma}\biggr)
  \Biggr]V^{\beta}\eta^{\gamma}V^{\delta}\\ - \frac{1}{6f'(R)}\biggl(\kappa T + 2f(R,T) -3\square f'(R)+f'(T)(T+4p)\biggr)\bigl(\delta_{\gamma}^{\alpha}g_{\delta\beta} - \delta_{\delta}^{\alpha}g_{\gamma\beta}\bigr)V^{\beta}\eta^{\gamma}V^{\delta}.
 \label{17}
\end{multline}
In the next section, we will use these results to obtain the GDE in GR equivalent
of $f(R,T)$ gravity model using FLRW metric. Naturally our results of this model will be acceptable
provided that the $f(R,T)$ limit case.

\subsection{Geodesic Deviation Equation for the FLRW universe}
For homogeneous and isotropic spacetimes, the Weyl tensor  $C_{\alpha\beta\gamma\delta}$ is identically zero and by using the FLRW metric (\ref{10}), the Ricci tensor (\ref{7c}) and the Ricci scalar
(\ref{7d}) becomes
\begin{equation}
R_{\alpha\beta}  = \frac{1}{f'(R)}\biggl[\bigl(\kappa+f'(T)\bigr)(\rho + p)u_{\alpha}u_{\beta}+ \biggl(\kappa p +2pf'(T)+ \frac{f(R,T)}{2}\biggr)g_{\alpha\beta} +
\mathcal{D}_{\alpha\beta}f'(R)\biggr],\label{18}
\end{equation}
\begin{equation}
R  = \frac{1}{f'(R)}\biggl[\kappa (3p-\rho) + 2f(R,T) - 3\square f'(R)+f'(T)(7p-\rho) \biggr].\label{19}
\end{equation}
From the known expressions of the Ricci tensor and the Ricci scalar, we can evaluate the Riemann tensor as following form
\begin{multline}
R_{\alpha\beta\gamma\delta} = \frac{1}{2f'(R)}\Biggl[\bigl(\kappa+f'(T)\bigr)(\rho+p)\bigl(u_{\delta}u_{\beta}g_{\alpha\gamma}-u_{\gamma}u_{\beta}g_{\alpha\delta} + u_{\gamma}u_{\alpha}g_{\beta\delta}-u_{\delta}u_{\alpha}g_{\beta\gamma}\bigr)\\
+ \biggl(\kappa p + \frac{\kappa\rho}{3} + \frac{f(R,T)}{3}+ \square f'(R)-\frac{1}{3}f'(T)(7p-\rho) \biggr)\bigl(g_{\alpha\gamma}g_{\delta\beta} - g_{\alpha \delta}g_{\gamma\beta}\bigr)+ (g_{\alpha\gamma}\mathcal{D}_{\delta\beta}
 - g_{\alpha\delta}\mathcal{D}_{\gamma\beta} + g_{\beta\delta}\mathcal{D}_{\gamma\alpha}- g_{\beta\gamma}\mathcal{D}_{\delta\alpha})f'(R)\\+
 2pf'(T)\biggl(g_{\delta\beta}g_{\alpha\gamma}-g_{\gamma\beta}g_{\alpha\delta}+g_{\gamma\alpha}g_{\beta\delta}-g_{\delta\alpha}g_{\beta\gamma}\biggr)\Biggr].
 \label{20}
\end{multline}
The vector field normalization implies that  $V^{\alpha}V_{\alpha} = \epsilon$, and we have
\begin{multline}
R_{\alpha\beta\gamma\delta}V^{\beta}V^{\delta} = \frac{1}{2f'(R)}\Biggl[\bigl(\kappa+f'(T)\bigr)(\rho+p)\bigl(g_{\alpha\gamma}(u_{\beta}V^{\beta})^2-2(u_{\beta}V^{\beta})V_{(\alpha}u_{\gamma)}
 + \epsilon u_{\alpha}u_{\gamma}\bigr)\\ + \biggl(\kappa p + \frac{\kappa\rho}{3} + \frac{f(R,T)}{3}+ \square f'(R)+\frac{1}{3}f'(T)(5p+\rho)\biggr)\bigl(\epsilon g_{\alpha\gamma}-V_{\alpha}V_{\gamma}\bigr)
 + (g_{\alpha\gamma}\mathcal{D}_{\delta\beta} - g_{\alpha\delta}\mathcal{D}_{\gamma\beta} + g_{\beta\delta}\mathcal{D}_{\gamma\alpha}- g_{\beta\gamma}\mathcal{D}_{\delta\alpha})f'(R)V^{\beta}V^{\delta}\Biggr].
 \label{21}
\end{multline}
 Contracting the last expression by $\eta^{\gamma}$ after rising its first index, one gets
\begin{multline}
R_{\beta\gamma\delta}^{\alpha}V^{\beta}\eta^{\gamma}V^{\delta} = \frac{1}{2f'(R)}\Biggr[\bigl(\kappa+f'(T)\bigr)(\rho+p)\bigl((u_{\beta}V^{\beta})^2\eta^{\alpha}-(u_{\beta}V^{\beta})V^{\alpha}(u_{\gamma}\eta^{\gamma})-(u_{\beta}V^{\beta})u^{\alpha}(V_{\gamma}\eta^{\gamma}) + \epsilon u^{\alpha}u_{\gamma}\eta^{\gamma}\bigr)\\
+ \biggl(\kappa p + \frac{\kappa\rho}{3} + \frac{f(R,T)}{3}+ \square f'(R)+\frac{1}{3}f'(T)(5p+\rho) \biggr)\bigl(\epsilon \eta^{\alpha}-V^{\alpha}(V_{\gamma}\eta^{\gamma})\bigr) + \bigl[(\delta_{\gamma}^{\alpha}\mathcal{D}_{\delta\beta} - \delta_{\delta}^{\alpha}\mathcal{D}_{\gamma\beta} +
g_{\beta\delta}\mathcal{D}_{\gamma}^{\, \, \alpha}- g_{\beta\gamma}\mathcal{D}_{\delta}^{\, \,  \alpha})f'(R)\bigr]V^{\beta}V^{\delta}\eta^{\gamma}\Biggr].
\label{22}
\end{multline}
Using $E = - V_{\alpha}u^{\alpha}$ and $\eta_{\alpha}u^{\alpha}=\eta_{\alpha}V^{\alpha}=0$ \cite{eti7}, Eq.(\ref{22}) reduces to
\begin{multline}
R_{\beta\gamma\delta}^{\alpha}V^{\beta}\eta^{\gamma}V^{\delta} = \frac{1}{2f'(R)}\Biggr[\bigl(\kappa+f'(T)\bigr)(\rho+p)E^2   + \epsilon\biggl(\kappa p + \frac{\kappa\rho}{3} + \frac{f(R,T)}{3}+ \square f'(R)+\frac{1}{3}f'(T)(5p+\rho)\biggr)\Biggr]\eta^{\alpha}\\
+ \frac{1}{2f'(R)}\biggl[\bigl[(\delta_{\gamma}^{\alpha}\mathcal{D}_{\delta\beta} - \delta_{\delta}^{\alpha}\mathcal{D}_{\gamma\beta} +
g_{\beta\delta}\mathcal{D}_{\gamma}^{\, \, \alpha}- g_{\beta\gamma}\mathcal{D}_{\delta}^{\, \,  \alpha})f'(R)\bigr]V^{\beta}V^{\delta}\biggr]\eta^{\gamma}.
\label{23}
\end{multline}
In the FLRW case, the covariant derivatives are
\begin{eqnarray}
\nabla_{0}\nabla_{0}f'(R) =f''(R)\ddot{R}+f'''(R) \dot{R}^{2}, \nabla_{i}\nabla_{j}f'(R)=-H g_{ij}f''(R)\dot{R},\quad  \square f'(R) = -f''(R)(\ddot{R}+3H\dot{R})-f'''(R) \dot{R}^2. 
\label{23a}
\end{eqnarray}
where we have used the definition for the Hubble parameter $H \equiv \frac{\dot{a}}{a}$ and $\dot{R} = \partial_{0} R$. 
Within the definition of the operator (\ref{7bb}) and using these results (\ref{23a}), the last term of the expression (\ref{23}) gives after cumbersome calculations
\begin{equation}
\bigl(\delta_{\gamma}^{\alpha}\mathcal{D}_{\delta\beta} - \delta_{\delta}^{\alpha}\mathcal{D}_{\gamma\beta} + g_{\beta\delta}\mathcal{D}_{\gamma}^{\, \, \alpha}
- g_{\beta\gamma}\mathcal{D}_{\delta}^{\, \,  \alpha}\bigr)f'(R)V^{\beta}V^{\delta}\eta^{\gamma} = \epsilon \bigl(5H f''(R) \dot{R}  + f''(R) \ddot{R} + f'''(R) \dot{R}^2\bigr) \eta^{\alpha}.
\label{24}
\end{equation}
Consequently  $R_{\beta\gamma\delta}^{\alpha}V^{\beta}\eta^{\gamma}V^{\delta}$ reduces to
\begin{equation}
R_{\beta\gamma\delta}^{\alpha}V^{\beta}\eta^{\gamma}V^{\delta} = \frac{1}{2f'(R)}\Biggl[\bigl(\kappa+f'(T)\bigr)(\rho+p)E^2 + \epsilon\biggl(\kappa p + \frac{\kappa\rho}{3} + \frac{f(R,T)}{3} +\frac{1}{3}f'(T)(5p+\rho)+ 2 H  f''(R) \dot{R}\biggr)\Biggr]\eta^{\alpha}.
\label{25}
\end{equation}
This equation is the generalization of the Pirani equation for the $f(R,T)$ metric formalism. Note that when $ f(R,T)= R-2\Lambda$ ie GR case with cosmological constant ($\Lambda$), Eq.(\ref{25}) leads to (\ref{15}).\\
Finally, we can write the GDE in $f(R,T)$ gravity as the following form 
\begin{equation}
\frac{D^2 \eta^{\alpha}}{D \nu^2} = - \frac{1}{2f'(R)}\Biggl[\bigl(\kappa+f'(T)\bigr)(\rho+p)E^2   + \epsilon\biggl(\kappa p + \frac{\kappa\rho}{3}+ \frac{f(R,T)}{3}+ \frac{1}{3}f'(T)(5p+\rho)+ 2 H  f''(R) \dot{R}\biggr)\Biggr]\eta^{\alpha}.
\label{26}
\end{equation}
We observe that the GDE in the FLRW case induces only a change in the magnitude of the deviation vector $\eta^{\alpha}$ 
and reflects spatial isotropy of space-time. Whereas in anisotropic universes, like Bianchi I,
the GDE also induces a change in the direction of the deviation vector, as indicated in \cite{eti8}

\subsection{GDE for fundamental observers with FLRW background}
 For the description case of the GDE for fundamental observes with FLRW background, we interpret $V^{\alpha}$ as the four-velocity
 of the fluid $u^{\alpha}$ and the the affine parameter $\nu$ coincides with the proper time of the central fundamental observer,
 ie $\nu =t$. with $\epsilon=-1$ (timelike geodesics) and also the vector field are normalized $E = 1$ , thus from (\ref{25}) one gets,
\begin{equation}
R_{\beta\gamma\delta}^{\alpha}u^{\beta}\eta^{\gamma}u^{\delta} = \frac{1}{2f'(R)}\biggl[\frac{2}{3}\rho\bigl(\kappa+f'(T)\bigr) -\frac{2}{3}pf'(T) -\frac{f(R,T)}{3} - 2 H  f''(R) \dot{R}\biggr]\eta^{\alpha}.
\label{27}
\end{equation}
Let the deviation vector be $\eta_{\alpha} = \ell e_{\alpha}$ where $ e_{\alpha}$ is parallel propagated along $t$,
 isotropy implies
\begin{equation}
\frac{D e^{\alpha}}{D t} = 0,
\end{equation}
and
\begin{equation}
\frac{D^2 \eta^{\alpha}}{D t^2} = \frac{d^2\ell}{dt^2} e^{\alpha}.
\end{equation}
Putting these results in the GDE (\ref{12}) and (\ref{27}), on gets
\begin{equation}
\frac{d^2\ell}{dt^2} = - \frac{1}{2f'(R)}\biggl[\frac{2}{3}\rho \bigl(\kappa+f'(T)\bigr)  -\frac{f(R,T)}{3}-\frac{2}{3}pf'(T) - 2 H  f''(R) \dot{R}\biggr]\, \ell .
\label{28}
\end{equation}
If we consider the particular case where $\ell = a(t)$, we have
\begin{equation}
\frac{\ddot{a}}{a} = \frac{1}{f'(R)}\biggl[\frac{f(R,T)}{6}+\frac{1}{3}pf'(T)  + H f''(R) \dot{R} -\frac{1}{3}\rho\bigl(\kappa+f'(T)\bigr) \biggr].
\label{29}
\end{equation}
This equation is a particular case of the generalized Raychaudhuri equation given in \cite{eti9}.
We mention here that with the standard forms of the modified Friedmann equations in $f(R,T)$  gravity model for flat universe \cite{eti10},
the above generalized Raychaudhuri equation can be obtained. These equations are written as follows

\begin{equation}
H^2 + \frac{k}{a^2}  = \frac{1}{3f'(R)}\biggl[\kappa\rho + \frac{\bigl(R f'(R) - f(R,T)\bigr)}{2} +f'(T)(\rho+p)  - 3H f''(R) \dot{R} \biggr],
\label{30}
\end{equation}
and
\begin{equation}
2\dot{H} + 3H^2 + \frac{k}{a^2}  = -\frac{1}{f'(R)}\biggl[\kappa p + 2H f''(R) \dot{R} + \frac{\bigl(f(R,T)- R f'(R)\bigr)}{2} -pf'(T) +f''(R) \ddot{R} + f'''(R) \dot{R}^2\biggr].
\label{31}
\end{equation}

\subsection{GDE for nulll vector fields with FLRW background}
Let us now restrict our investigation to the GDE for null vector
fields past directed , in this case $V^{\alpha}=k^{\alpha}$ with $k_{\alpha}k^{\alpha}=0$ and consequently $\epsilon = 0$.
The generalized pirani Equation (\ref{25}) then reduces to
\begin{equation}
R_{\beta\gamma\delta}^{\alpha}k^{\beta}\eta^{\gamma}k^{\delta} = \frac{1}{2f'(R)}\bigl(\kappa+f'(T)\bigr)(\rho+p)E^2\,\eta^{\alpha}.
\label{32}
\end{equation}
This equation can be interpreted as the \textit{Ricci focusing} in $f(R,T)$ gravity.
Let us now consider $\eta^{\alpha}= \eta e^{\alpha}$,  $e_{\alpha}e^{\alpha}=1$, $e_{\alpha}u^{\alpha}=e_{\alpha}k^{\alpha}=0$ 
and using a basis which is both parallel propagated and aligned, ie
$\frac{D e^{\alpha}}{D \nu}=k^{\beta}\nabla_{\beta}e^{\alpha}=0$, we obtain from the Eq. (\ref{26}), the GDE for null vector
as the following form
\begin{equation}
\frac{d^2\eta}{d\nu^2} = - \frac{1}{2f'(R)}\bigl(\kappa+f'(T)\bigr) (\rho+p)E^2\, \eta,
\label{33}
\end{equation}
which expresses the focusing of all families of past directed geodesics provided that ($\rho_{total}+p_{total})>0$ is satisfied.
At this stage the usual GR result discussed in \cite{eti7} is recovered if  we have $\kappa(\rho+p) > 0$, and
that a cosmological constant term in the gravitational Lagrangian with equation of state $p = - \rho$ 
does not affect the focusing of null of null geodesics \cite{eti22}.
In the same way we deduct from Eq.(\ref{33}), the focusing condition for $f(R,T)$ gravity given by
\begin{equation}
\frac{\bigl(\kappa+f'(T)\bigr)(\rho + p)}{f'(R)} > 0.
\end{equation}
 Eq.(\ref{33}) can be evaluated in function of the redshift parameter $z$. To do so, we write 
\begin{equation}
\frac{d}{d\nu} = \frac{dz}{d\nu}\frac{d}{dz},
\end{equation}
which results in
\begin{align}
\frac{d^2}{d\nu^2} &= \frac{dz}{d\nu}\frac{d}{dz}\biggl(\frac{d}{d\nu}\biggr), \notag \\
& = \biggl(\frac{d\nu}{dz}\biggr)^{-2}\biggl[-\biggl(\frac{d\nu}{dz}\biggr)^{-1}\frac{d^2\nu}{dz^2}\frac{d}{dz}+\frac{d^2}{dz^2}\biggr].
\label{33a}
\end{align}
Let us consider the null geodesics for which we have
\begin{equation}
(1+z) = \frac{a_0}{a}=\frac{E}{E_0} \quad \longrightarrow \quad \frac{dz}{1+z}=- \frac{da}{a}.
\end{equation}
By choosing the present value of the scalar factor ($a_0=1$), we gets for the past directed case
\begin{equation}
dz = (1+z) \frac{1}{a}\frac{da}{d\nu}\, d\nu = (1+z) \frac{\dot{a}}{a} E\, d\nu = E_0 H (1+z)^2\, d\nu,
\end{equation}
which leads to
\begin{equation}
\frac{d\nu}{dz} = \frac{1}{E_0 H (1+z)^2}.
\end{equation}
The second derivative of this expression gives
\begin{equation}
\frac{d^2\nu}{dz^2} =- \frac{1}{E_0 H (1+z)^3}\biggl[\frac{1}{H}(1+z)\frac{dH}{dz}+2\biggr],
\end{equation}
where
\begin{equation}
\frac{dH}{dz} = \frac{d\nu}{dz}\frac{dt}{d\nu}\frac{dH}{dt} = - \frac{1}{H(1+z)} \frac{dH}{dt},
\end{equation}
and we notice that we also used $\frac{dt}{d\nu} = E_0 (1+z)$.
From the definition of the Hubble parameter $H= \frac{\dot{a}}{a}$, on gets
\begin{equation}
\dot{H} = \frac{\ddot{a}}{a} - H^2,
\label{34}
\end{equation}
Putting the expression (\ref{29}) in (\ref{34}), we have
\begin{equation}
\dot{H} = \frac{1}{f'(R)}\biggl[\frac{f(R,T)}{6}  + H f''(R) \dot{R} -\frac{\kappa \rho}{3}  +\frac{1}{3}f'(T) (p-\rho) \biggr]- H^2,
\end{equation}
then
\begin{equation}
\frac{d^2\nu}{dz^2} = -\frac{3}{E_0 H (1+z)^3}\biggl[1+ \frac{1}{3H^2 f'(R)}\biggl(\frac{\kappa \rho}{3}- \frac{f(R,T)}{6}-\frac{f'(T)}{3}(p-\rho) - H f''(R) \dot{R}\biggr)\biggr].
\label{35a}
\end{equation}
By substituting Eq.(\ref{35a}) in (\ref{33a}), the operator $\frac{d^2\eta}{d\nu^2}$ leads to
\begin{equation}
\frac{d^2\eta}{d\nu^2} = \bigl(EH(1+z)\bigr)^2\Biggl[\frac{d^2\eta}{dz^2} + \frac{3}{(1+z)}\biggl[1+ \frac{1}{3H^2 f'(R)}\biggl
(\frac{\kappa \rho}{3}- \frac{f(R,T)}{6}-\frac{f'(T)}{3}(p-\rho) - H f''(R) \dot{R}\biggr)\biggr]\frac{d\eta}{dz}\Biggr].
\label{35}
\end{equation}
Finally the GDE (\ref{33}) for null vector fields in $f(R,T)$ gravity takes the following form
\begin{equation}
\frac{d^2\eta}{dz^2} + \frac{3}{(1+z)}\Biggl[1+ \frac{1}{3H^2 f'(R)}\biggl
(\frac{\kappa \rho}{3}- \frac{f(R,T)}{6} -\frac{f'(T)}{3} (p-\rho)- H f''(R) \dot{R}\biggr)\Biggr]\, \frac{d\eta}{dz} + \frac{\bigl(\kappa+f'(T)\bigr)(\rho+p)}{2H^2(1+z)^2f'(R)}\, \eta = 0.
\label{36}
\end{equation}
If we suppose that the content of the universe is the ordinary matter (dust) and the radiation, 
the energy density $\rho$ and the pressure $p$  could be written as
\begin{equation}
\kappa\rho = 3H_0^2 \Omega_{m0}(1 + z)^3 + 3H_0^2\Omega_{r0}(1 + z)^4, \qquad \kappa p = H_0^2\Omega_{r0}(1 + z)^4,
\label{37}
\end{equation}
where we have used $p_m=0$ and $p_r = \frac{1}{3}\rho_r$.
From the Eq.(\ref{30}), $H^2$ gives
\begin{eqnarray}
H^2 &=& \frac{1}{f'(R)}\biggl[H_0^2 \Omega_{m0}(1 + z)^3 \bigr(1+f'(T)\bigr) + H_0^2\Omega_{r0}(1 + z)^4 \bigl(1+\frac{4f'(T)}{3}\bigr) + \frac{\bigl(R f'(R)-f(R,T)\bigr)}{6} - H f''(R) \dot{R}\biggr] - \frac{k}{a^2}, \notag \\
&=& \frac{H_0^2}{f'(R)}\biggl[ \Omega_{m0}(1 + z)^3 + \Omega_{r0}(1 + z)^4 + \Omega_{DE} + \Omega_{k}(1+z)^2 f'(R)\biggr],
\label{37a}
\end{eqnarray}
where  the Dark Energy parameter $\Omega_{DE}$ is
\begin{equation}
\Omega_{DE} = \frac{1}{H_0^2 }\biggl[\frac{\bigl(R f'(R)-f(R,T)\bigr)}{6} - H f''(R) \dot{R} +{H_0} ^{2}f'(T)\bigl(\Omega_{m0}(1 + z)^3 +\frac{4}{3} \Omega_{r0}(1 + z)^4 \bigr) \biggr].
\label{37b}
\end{equation}
By using the Eqs.(\ref{37}) and (\ref{37a}), the GDE for null vector field takes the following form
\begin{equation}
\frac{d^2\eta}{dz^2} + \mathcal{P}_1(H,R,z)\frac{d\eta}{dz} + \mathcal{Q}_1(H,R,z)\eta = 0,
\label{38}
\end{equation}
where
\begin{equation}
 \mathcal{P}_1(H,R,z) = \frac{4\Omega_{m0}(1 + z)^3 + 4\Omega_{r0}(1 + z)^4 +  3f'(R)\Omega_{k0}(1 + z)^2  + 4\Omega_{DE}-\frac{10}{9}\Omega_{r0}{(1+z)}^{4}f'(T) -\frac{Rf'(R)}{6 H_0^2}}{(1+z)\bigl(\Omega_{m0}(1 + z)^3 + \Omega_{r0}(1 + z)^4 + f'(R)\Omega_{k0}(1 + z)^2 + \Omega_{DE}\bigr)},
\label{39}
\end{equation}
\begin{equation}
\mathcal{Q}_1(H,R,z) = \frac{\biggl(3\Omega_{m0}(1 + z) + 4\Omega_{r0}(1 + z)^2 \biggr)\bigl(1+f'(T)\bigr) }{2\bigl(\Omega_{m0}(1 + z)^3+\Omega_{r0}(1 + z)^4 +  f'(R)\Omega_{k0}(1 + z)^2 + \Omega_{DE}\bigr)}.
\label{40}
\end{equation}
and
\begin{equation}
\Omega_{k0}=-\frac{k}{H_0^2 a_0^2}.
\end{equation}
Eq.(\ref{38}) is difficult to solve analytically. In order to solve its, it is possible to rewrite
as a differential equation containing only the unknown $H(Z)$ and $f(z)$  and their derivatives with respect to $z$.
This method have been widely developed in \cite{eti11}-\cite{eti12}.
For this we fix $H=H(z)$, and we express the Ricci scalar $R$ and the trace $T$ of energy momentum tensor as
\begin{eqnarray}
R=6\biggl[2H^2  -(1+z)H \frac{dH}{dz} + k(1+z)^2 \biggr],
\label{41}
\end{eqnarray}
\begin{eqnarray}
T &=& 3p-\rho  \cr
&=& \frac{3H_0 ^2}{\kappa}\Omega_{m0}{(1+z)}^{3}.
\label{42}
\end{eqnarray}
To examine these results with GR, we may consider the limit case $f(R,T) = R-2\Lambda$. For this choice we have,
$f'(R) = 1$, $f''(R) =0$, $f(T) = -2\Lambda$ , $f'(T) = 0$. Then the expression for $\Omega_{DE}$ reduces to
\begin{equation}
\Omega_{DE} = \frac{1}{H_0^2 }\biggl[\frac{(R-R + 2\Lambda)}{6}\biggr] = \frac{\Lambda}{3H_0^2} \equiv \Omega_{\Lambda}.
\end{equation}
The first Friedmann modified equation (\ref{37a}) reduces to the well known expression in GR.
\begin{equation}
H^2 = H_0^2\bigl[\Omega_{m0}(1 + z)^3 + \Omega_{r0}(1 + z)^4 + \Omega_{\Lambda}  + \Omega_{k}(1+z)^2\bigr],
\end{equation}
 and the expressions $\mathcal{P}_1$ (\ref{39}), $\mathcal{Q}_1$ (\ref{40}) takes the following forms
\begin{equation}
\mathcal{P}_1(z) = \frac{4\Omega_{r0}(1 + z)^4 + (7/2)
\Omega_{m0}(1 + z)^3 + 3\Omega_{k0}(1 + z)^2 + 2\Omega_{\Lambda}
}{(1+z)\bigl(\Omega_{r0}(1 + z)^4 + \Omega_{m0}(1 + z)^3 + \Omega_{k0}(1 + z)^2 + \Omega_{\Lambda}\bigr)},
\end{equation}
\begin{equation}
\mathcal{Q}_1(z) = \frac{2\Omega_{r0}(1 + z)^2 + (3/2)
\Omega_{m0}(1 + z)}{\Omega_{r0}(1 + z)^4 + \Omega_{m0}(1 + z)^3 + \Omega_{k0}(1 + z)^2 + \Omega_{\Lambda}}.
\end{equation}
Then, in GR the GDE for null vector fields with FLRW background is 
\begin{multline}
\frac{d^2\eta}{dz^2} + \frac{4\Omega_{r0}(1 + z)^4 + (7/2)
\Omega_{m0}(1 + z)^3 + 3\Omega_{k0}(1 + z)^2 + 2\Omega_{\Lambda}
}{(1+z)\bigl(\Omega_{r0}(1 + z)^4 + \Omega_{m0}(1 + z)^3 + \Omega_{k0}(1 + z)^2 + \Omega_{\Lambda}\bigr)}\, \frac{d\eta}{dz}\\ + \frac{2\Omega_{r0}(1 + z)^2 + (3/2)
\Omega_{m0}(1 + z)}{\Omega_{r0}(1 + z)^4 + \Omega_{m0}(1 + z)^3 + \Omega_{k0}(1 + z)^2 + \Omega_{\Lambda}}\, \eta = 0.
\end{multline}
Notice that when we fix $\Omega_{\Lambda} = 0$ which leads to $\Omega_{m0} + \Omega_{r0}+ \Omega_{k0}=1$, we obtain the mattig relation in GR given by \cite{eti13}
\begin{equation}
\frac{d^2\eta}{dz^2} + \frac{6 +
\Omega_{m0}(1 + 7z) + \Omega_{r0}(1 + 8z + 4z^2)}{2(1 + z)(1 + \Omega_{m0}z +
\Omega_{r0}z(2 + z))}\, \frac{d\eta}{dz} + \frac{3\Omega_{m0} + 4\Omega_{r0}(1 + z)}{2(1 + z)(1 + \Omega_{m0}z +
\Omega_{r0}z(2 + z))}\, \eta = 0.
\end{equation}
Equation (\ref{38}) could be interpreted as a generalization of the Mattig relation in $f(R,T)$ gravity.\\
Within the explicit form for the deviation vector of a (past-directed) 
geodesic null congruence given by this equation, we are now in a position to easily infer an expression for the
observer area distance $r_0 (z)$ \cite{eti13},
\begin{eqnarray}
r_0(z) = \sqrt{ \Biggl\rvert \frac{dA_0(z)}{d\Omega} \Biggr\rvert} = \Biggl\rvert \frac{\eta(z')\mid z'=z}{d\eta(z')/dl \mid z'= 0}\Biggr\rvert\cr
\end{eqnarray}
where $A_0$ is the area of the object and also $\Omega$ is the solid angle. We have used the fact that $d/dl = {E_0}^{-1} {(1 + z)}^{-1} d/d\nu =
H(1 + z)d/dz$ and choosing the deviation to be zero at $z=0$. Thus we gets

\begin{eqnarray}
r_0(z) = \Biggl\rvert \frac{\eta(z)}{ H(0)d\eta(z')/dz' \mid z'= 0}\Biggr\rvert\ ,
\label{43}
\end{eqnarray}
being $H(0)$  the modified Friedmann equation (\ref{37a}) at $z = 0$.
Analytical expression for the observable area distance for GR with no cosmological constant can be found in
\cite{eti22, eti26,eti27}, whereas for more general scenarios numerical
integration is usually required.
In the next section, the numerical analysis also will be performed in order to find the deviation vector $\eta(z)$ and observer area distance
 $r_0 (z)$ and compare the results have been found in $ f(R,T)$ Jordan frame with those $\Lambda CDM$ model. 

\subsection{ Numerically solution of GDE for null vector fields in $f(R,T)$ gravity}
To solve numerically the null vector GDE  in $f(R,T)$ gravity, we concentrate one particular model of $f(R,T)$ gravity,
namely $R+f(T)$, where  $ f(T)= \alpha_1 T^{\beta_1}$; $\alpha_1$ and $\beta_1 = \frac{1+3w}{2(1+w)}$  being constant. The dynamics and stability of this model are studied  
 and interesting results have been found in \cite {eti2}. Thus for this model, the equations (\ref{37b}), (\ref{39}, (\ref{40}) can be rewritten
 as follow
\begin{eqnarray}
\Omega_{DE} = \frac{1}{H_0^2 }\biggl[-\frac{\alpha_1 T^{\beta_1}}{6}+ {H_0} ^{2}\alpha_1 \beta_1 T^{\beta_1-1} \bigl(\Omega_{m0}(1 + z)^3 +\frac{4}{3} \Omega_{r0}(1 + z)^4 \bigr) \biggr],
\end{eqnarray}
\begin{eqnarray}
\mathcal{P}_1(H,R,z) = \frac{4\Omega_{m0}(1 + z)^3 + 4\Omega_{r0}(1 + z)^4 +  3\Omega_{k0}(1 + z)^2  + 4\Omega_{DE}-\frac{10}{9}\alpha_1\beta_1\Omega_{r0}{(1+z)}^{4} T^{(\beta_1-1)} -\frac{R}{6 H_0^2}}{(1+z)\bigl(\Omega_{m0}(1 + z)^3 + \Omega_{r0}(1 + z)^4 + \Omega_{k0}(1 + z)^2 + \Omega_{DE}\bigr)}, 
\end{eqnarray}
\begin{eqnarray}
\mathcal{Q}_1(H,R,z) = \frac{\biggl(3\Omega_{m0}(1 + z) + 4\Omega_{r0}(1 + z)^2 \biggr)\bigl(1+\alpha_1\beta_1 T^{(\beta_1-1)}\bigr) }{2\bigl(\Omega_{m0}(1 + z)^3+\Omega_{r0}(1 + z)^4 +  \Omega_{k0}(1 + z)^2 + \Omega_{DE}\bigr)}.
\end{eqnarray}

 \begin{figure}[h]
  \centering
  \begin{tabular}{rl}
  \includegraphics[width=9cm, height=9cm]{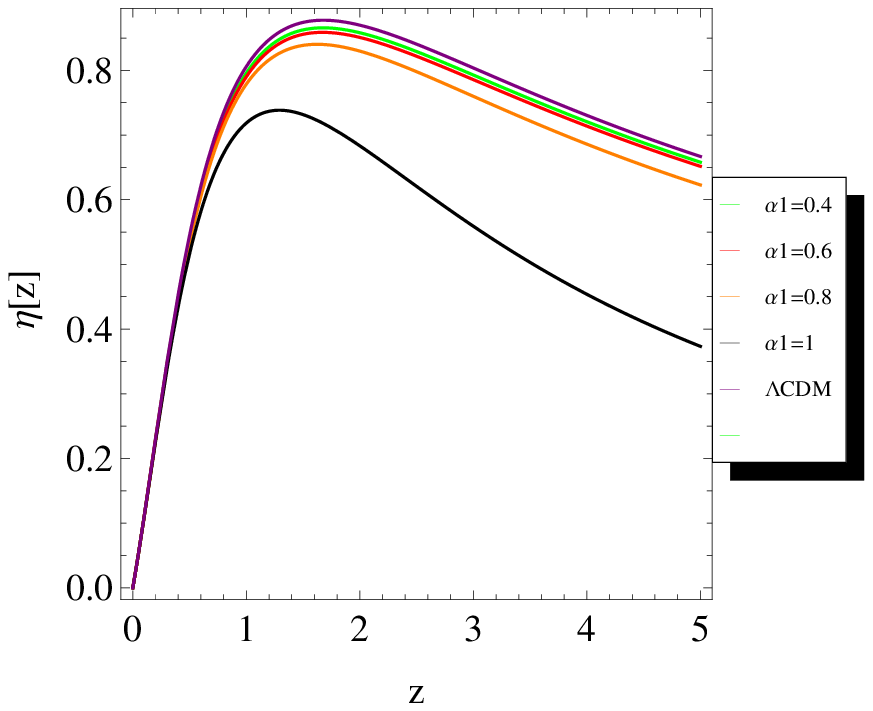}&
  \includegraphics[width=9cm, height=9cm]{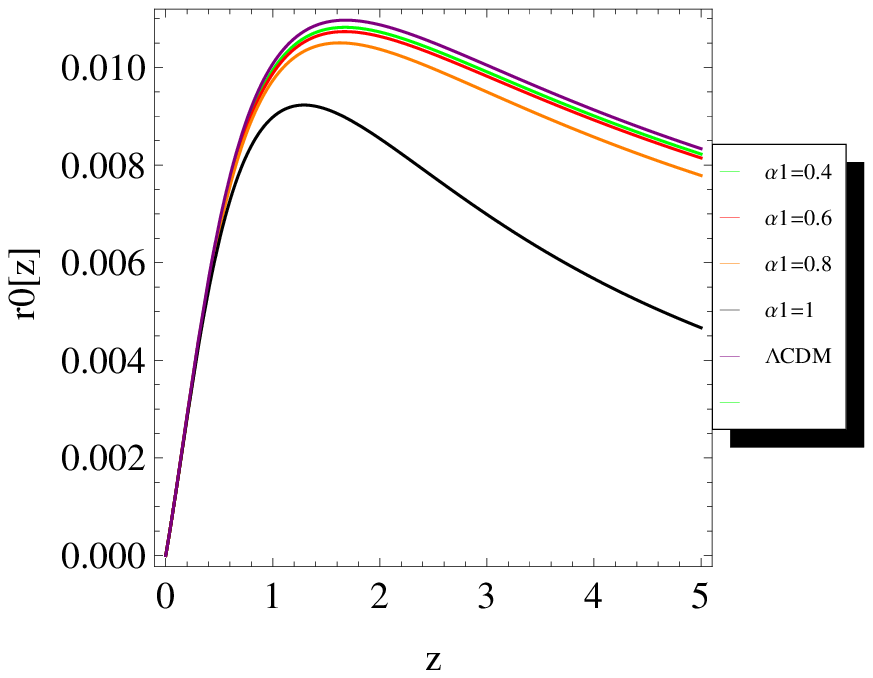}
  \end{tabular}
  \caption{The graphs shows the deviation vector magnitude $\eta(z)$ (left panel) and observer area distance $r_0(z)$ (right panel) for null
  vector field GDE with FLRW background as functions of redshift. The graphs are plotted for $H_0 = 80Km/s/Mpc$, $\Omega_{m0}=0.3$, $\Omega_{r0}=\Omega_{k0}=0$, $\Lambda=1.7.10^{-121}$ 
  and we imposed in equation (\ref{38}) the initial conditions $\eta(z=0)= 0$ and $\eta'(z=0)=1$.}
 \label{fig2}
  \end{figure}
In each panel of fig2, we observe that the evolution of the deviation $\eta(z)$ and  observer area distance  $r_0(z)$ are similar behavior to those of $\Lambda CDM$. Within
the model $f(R,T)= R+ \alpha_1 T^{\beta_1}$, we see as the $\alpha_1$  increases ($\alpha_1 \geqslant 1$) and when one goes to the highest values of the redshifts $(z\geqslant 0.8)$, the 
deviation $\eta(z)$ and observer area distance  $r_0(z)$ decouple each of the model $\Lambda CDM$ but still keeps the same pace,
while for the low values of the redshifts ie for the current day, the $f(R,T)$ model reproduce exactely $\Lambda CDM$.
We can conclude that for all the cases considered the results are similar to $\Lambda CDM$, which means that the above model $f(R,T)$ considered 
remain phenomenologically viable and can be tested with observational data.

%%%%%%%%%%%%%%%%%%%%%%%%%%%%%%%%%%%%%%%%%%%%%%%%%%%%%%%%%%%%%%%%%%%%%%%%%%%%%%%%%%%%%%%%%%%%%%%%%%%%%%%%%%%%%%%%%%%%%%%
\section{Conclusion}
In this paper, we have presented the Geodesic Deviation Equation (GDE) in the metric formalism of $f(R,T)$ theories.
The calculations of the Ricci scalar and the Riemann tensor have been done firtly by using the $f(R,T)$ fields gravity equations.
The Geodesic Deviation Equation and the generalization of the pirani equation for the FLRW universe in $f(R,T)$ gravity is investigated
and both these equations reduces to the well known relation when $f(R,T) = R-2\Lambda$. We focused  our attention on two 
particular cases, the GDE for fundamental observes and the past directed null vector fields with FLRW universe. Within these cases
we have obtained the Raychaudhuri equation, the generalized Mattig relation and the diametric angular distance
differential for $f(R,T)$ gravity theory. Furthermore, in the similar way to the GR, the focusing for past-directed null geodesics condition for $f(R,T)$ gravity is be done.
Numerically results of the geodesic deviation $\eta(z)$ and observer area distance $r_0(z)$ for the $f(R,T)$ model where compared
with  those of the  $\Lambda CDM$ model. We have also obtained the geodesic deviation $\eta(z)$ and the area distance $r0(z)$
corresponding to the $f(R,T)$ model and compared them with those of ΛCDM model.

%%%%%%%%%%%%%%%%%%%%%%%%%%%%%%%%%%%%%%%%%%%%%%%%%%%%%%%%%%%%%%%%%%%%%%%%%%%%%%%%%%%%%%%%%%%%%%%%%%%%%%%%%%%%%%%%%%%%%%%%%

\acknowledgments
 E. H. Baffou  thanks
IMSP and the Governement of Benin for every kind of support during the realization of this work. M. J. S. Houndjo and A. V. Kpadonou thank {\it Ecole Normale Sup\'erieure de Natitingou}.  M. E. Rodrigues also expresses his sincere gratitude to UFPA and CNPq.

\newpage
%%%%%%%%%%%%%%%%%%%%%%%%%%%%%%%%%%%%%%%%%%%%%%%%%%%%%%%%%%%%%%%%%%%%%%%%%%%%%%%%%%%%%%%%%%%%%%%%%%%%%%%%%%%%%%%%%%%%%%%%%%%%%


\begin{thebibliography}{17}
%%%%%%%%%%%%%%%%%%%%%%%%%%%%%%%%%%%%%%%%%%%%%%%%%%%%%%%%%%%%%%%%%%%%%%%%%%%%%%%%%%%%%%%%%%%%%%%%%%%%%%%%%%%%%%%%%%%%%%%%%%%%%%
\addcontentsline{toc}{chapter}{Bibliographie}


\bibitem{eti29} C. Corda. Interferometric detection of gravitational waves: the definitive test for General Relativity.
Int. J. Mod. Phys. D 18:2275-2282 (2009). Preprint in [arXiv:0905.2502].


\bibitem{eti25} Szekeres P: The Gravitational Compass, J. Maths. Phys. 6 (1965), 1387.


\bibitem{eti20} J. L. Synge. On the Deviation of Geodesics and Null Geodesics, Particularly in Relation to the
Properties of Spaces of Constant Curvature and Indefinite Line Element. Ann. Math. 35:705 (1934).


\bibitem{eti21}
F. A. E. Pirani. On the Physical Significance of the Riemann Tensor. Acta Phys. Polon. 15:389
(1956).

\bibitem{eti22} G. F. R. Ellis and H. Van Elst. Deviation of geodesics in FLRW spacetime geometries (1997). Preprint
in [arXiv:gr-qc/9709060v1].

\bibitem{eti23} S.L. Shapiro and S.A. Teukolsky, Black Holes, White Dwarfs and Neutron Satrs (Wile-Interscience, New York 1983).

\bibitem{eti24} K.S. Thorne, in S. Hawking and W. Israel, eds, 300 Year of
Gravitation (Cambridge University, Cambridge 1987) p. 330.


\bibitem{eti26} Raychaudhuri A K: Relativistic Cosmology, Phys. Rev. 98 (1955), 1123.

\bibitem{eti27} Mattig W: Uber den Zusammenhang zwischen Rotverschiebung und scheinbarer Helligkeit, Astr.
Nach. 284 (1958), 109.

\bibitem{eti28} Pirani F A E: On the Physical Significance of the Riemann Tensor, Acta Phys. Polon. 15 (1956),389.



\bibitem{eti15} A.G. Riess et al., Astron. J. 116, 1009 (1998); S. Perl- 
mutter, etet al., Nature 391, 51 (1998); S. Perlmutter 
etet al., Astrohpys. J. 517, 565 (1999). 

\bibitem{eti16} P.A.R.Ade et al. (Planck Collaboration), arXiv:1303.5062 [astro-ph.CO]; D.N. Spergel et al., 
Astrophys. J. Suppl. 170, 377S (2007).


\bibitem{eti17} S. Cole, W.J. Percival, J. A. Peacock, et al., Mon. Not. 
Roy. Astron. Soc. 362, 505 (2005); D.J. Eisenstein, I. 
Zehavi, D.W. Hogg, et al., ApJ 633, 560 (2005); W.J. 
Percival, B.A. Reid, D.J. Eisenstein, et al., Mon. Not. 
Roy. Astron. Soc. 401, 2148 (2010); N. Padmanabhan, 
X. Xu, D.J. Eisenstein, et al., Mon. Not. Roy. Astron. 
Soc. 427, 2132 (2012); C. Blake, E.A. Kazin, F. Beutler, 
et al, Mon. Not. Roy. Astron. Soc. 418, 1707 (2011); L. 
Anderson, E. Aubourg, S. Bailey, et al, Mon. Not. Roy. 
Astron. Soc. 428, 1036 (2013). 

\bibitem{eti18} Hawkins, E. et al.: Mon. Not. Roy. Astron. Soc. 346(2003)78; Tegmark,
M. et al.: Phys. Rev. D69(2004)103501; Cole, S. et al.: Mon. Not. Roy.
Astron. Soc. 362(2005)505.

\bibitem{eti19} Jain, B. and Taylor, A.: Phys. Rev. Lett. 91(2003)141302.

\bibitem{eti30}  H-J. Schmidt. Variational derivatives of arbitrarily high order and multi-inflation cosmological mod-
els. Class. Quantum. Grav. 7:1023-1031 (1990).

\bibitem{eti31} D. Wands. Extended gravity theories and the Einstein-Hilbert action. Class. Quant. Grav. 11:269-280
(1994). Preprint in [arXiv:gr-qc/9307034].

\bibitem{eti1} T. Harko, F. S. N. Lobo, S. Nojiri and S. D. Odintsov, Phys. Rev. D84(2011) 024020. [arXiv: 1104.2669 [gr-qc]].

\bibitem{eti32} M. J. S. Houndjo, Int. J. Mod. Phys. D. 21, 1250003 (2012). arXiv: 1107.3887 [astro-ph.CO]; M. J. S. Houndjo and O. F.
Piattella, Int. J. Mod. Phys. D. 21, 1250024 (2012). arXiv: 1111.4275 [gr.qc]; D. Momeni, M. Jamil and R. Myrzakulov,
Euro. Phys. J. C 72, arXiv: 1107.5807[physics.gen-ph].

\bibitem{eti33}
 M. Sharif and M. Zubair, JCAP 03, 028 (2012); arXiv:1204.0848v2 [gr-qc]. M. Jamil, D. Momeni and R. Myrzakulov, Chin.
Phys. Lett. 29, 109801 (2012) [arXiv:1209.2916 [physics.gen-ph]].

\bibitem{eti34}
M. J. S. Houndjo, C. E. M. Batista, J. P. Campos and O. F. Piattella, Can. J. Phys. 91, 548-553 (2013). arXiv:1203.6084
[gr-qc].

\bibitem{eti35}
A. Guarnizo, L. Castaneda, J. M. Tejeiro, Gen. Rel. Grav. 43, 2713 (2011);
A. de la Cruz-Dombriz, P. K. S. Dunsby, V. C. Busti, S. Kandhai, Phys. Rev. D 89, 064029 (2014), arXiv:1312.2022;
A. Guarnizo, L. Castaneda, J. M. Tejeiro, arXiv:1402.3196.

\bibitem{eti36}
F. Shojai and A. Shojai, Phys. Rev. D 78, 104011 (2008).

\bibitem{eti37} F. Darabi, M. Mousavi, K. Atazadeh,  Dec 31, 2014. 11 pp.
Published in Phys.Rev. D91 (2015) 084023.

\bibitem{eti2} E.H. Baffou, A.V. Kpadonou, M.E. Rodrigues, M.J.S. Houndjo, J. Tossa, Published in Astrophys.Space Sci. 355 (2014)
2197, arXiv:1312.7311 [gr-qc]


\bibitem{eti3} S. Capozziello, S. Carloni and A. Troisi. Quintessence without scalar fields. Recent Res. Dev. Astron.
Astrophys. 1, 625 (2003). Preprint in [arXiv:astro-ph/0303041].[arXiv:astro-ph/0303041]

\bibitem{eti4} R. M. Wald, General Relativity. The University of Chicago Press, 1984.

\bibitem{eti5}  E. Poisson, A Relativist’s Toolkit - The Mathematics of Black-Hole Mechanics. Cambridge University
Press, 2004

\bibitem{eti6} J. L. Synge, Ann. Math. 35, 705 (1934); F. A. E. Pirani, Acta Phys. Polon. 15, 389 (1956).


\bibitem{eti7}
G. F. R. Ellis and H. Van Elst, [arXiv:gr-qc/9812046v5].

\bibitem{eti8} D. L. Caceres, L. Casta ̃eda, J. M. Tejeiro. Geodesic Deviation Equation in Bianchi Cosmologies. J.
 Phys. Conf. Ser. 229:012076 (2010). Preprint in [arXiv:0912.4220v1].

\bibitem{eti9} S. Rippl, H. van Elst, R. Tavakol and D. Taylor. Kinematics and dynamics of f (R) theories of gravity.
Gen. Relativ. Gravit. 28:193 (1996).

\bibitem{eti10} E. H. Baffou , M. J. S. Houndjo, A. V. Kpadonou, M. E. Rodrigues, and J. Tossa, arXiv:1504.05496 [gr-qc].

\bibitem{eti11} S. Capozziello and M. Francaviglia. Extended theories of gravity and their cosmological and astro-
physical applications. Gen. Relativ. Gravit. 40:357-420 (2008). Preprint in [arXiv:0706.1146v2].

\bibitem{eti12} S. Capozziello, V. F. Cardone, and A. Troisi. Reconciling dark energy models with f (R) theories.
Phys. Rev. D. 71:043503 (2005).

\bibitem{eti13} P. Schneider, J. Ehlers and E. E. Falco, Gravitational Lenses, (Springer-Verlag, 1999).

\bibitem{eti14} A. Guarnizo, L. Castaneda, and J. M. Tejeiro, “Geodesic deviation equation in f (R) gravity,”
Gen. Rel. Grav. 43 (2011) 2713–2728, arXiv:1010.5279 [gr-qc].


\end{thebibliography}
\end{document}